\documentclass[journal=aamick,manuscript=article]{achemso}
\usepackage{natbib}
\usepackage{amsmath}
\usepackage{graphicx}
\usepackage{caption}
\usepackage{subcaption}
\usepackage[export]{adjustbox}
\usepackage{multirow}
\usepackage{hyperref}
\usepackage{booktabs}
\usepackage[T1]{fontenc}
\usepackage[utf8]{inputenc}

\usepackage{tablefootnote} 
\usepackage{threeparttable}
\hypersetup{colorlinks=true, urlcolor=blue, citecolor=blue}
\usepackage{chemmacros}

\author{Dimple Rani}
 \affiliation{School of Physical Sciences, National Institute of Science Education and Research,
An OCC of Homi Bhabha National Institute, Jatni 752050, India}
\email{dimple.rani@niser.ac.in}

\author{Subrata Jana}
\affiliation{Institute of Physics, Faculty of Physics, Astronomy and Informatics, Nicolaus Copernicus University in Toru\'n,
ul. Grudzi\k{a}dzka 5, 87-100 Toru\'n, Poland}

\author{Manish K Niranjan}
\affiliation{Department of Physics, Indian Institute of Technology, Hyderabad, India}
\author{Prasanjit Samal}
 \affiliation{School of Physical Sciences, National Institute of Science Education and Research,
An OCC of Homi Bhabha National Institute, Jatni 752050, India} 
\title[An \textsf{achemso} demo]
  { Exploring Thermoelectric Transport Properties of Ag-Based Chalcopyrite semiconductors via Dielectric-Dependent Hybrid DFT}

\abbreviations{IR,NMR,UV}

\begin{document}

\begin{tocentry}

Some journals require a graphical entry for the Table of Contents.
This should be laid out ``print ready'' so that the sizing of the
text is correct.

Inside the \texttt{tocentry} environment, the font used is Helvetica
8\,pt, as required by \emph{Journal of the American Chemical
Society}.

The surrounding frame is 9\,cm by 3.5\,cm, which is the maximum
permitted for  \emph{Journal of the American Chemical Society}
graphical table of content entries. The box will not resize if the
content is too big: instead it will overflow the edge of the box.

This box and the associated title will always be printed on a
separate page at the end of the document.

\end{tocentry}

\begin{abstract}
We explore the intricate domain of thermoelectric phenomena within silver (Ag) based chalcopyrites, focusing on compositions such as AgXTe$_2$ (where X=Ga, In) and the complex quaternary system Ag$_2$ZnSn/GeY$_2$ (with Y=S, Se). 
The thermoelectric parameters are estimated 
from the Boltzmann transport theory in combination of the electronic structures that, in turn, are obtained using the
non-empirical screened dielectric-dependent hybrid (DDH) functional approach within the density functional theory (DFT). This approach allows us to analyze critical thermoelectric properties, including electrical conductivity, Seebeck coefficient, and power factor. We investigate the critical role of phonon scattering by leveraging both the elastic constant tensor and the deformation potential method. This enables a rigorous examination of electron relaxation time and lattice thermal conductivity, enhancing the robustness of our predictions.  We identify materials with a thermoelectric Figure of merit (ZT = $\sigma S^{2}T/ \kappa$) exceeding the critical threshold of unity. 
 Our findings delineate a promising trajectory, laying the groundwork for the emergence of a new class of Ag-based chalcopyrites distinguished by their exceptional thermoelectric characteristics. 
\end{abstract}

\section{Introduction} 
Thermoelectric (TE) materials have captured significant interest, particularly in the last two decades, for their unique ability to directly convert heat into electricity without relying on moving components~\cite{disalvo1999thermoelectric,bell2008cooling}. 
 TE materials are positioned as crucial contributors to the evolution of a more sustainable global energy landscape,  offering promising applications, including onboard power for wearable electronic sensors and systems designed for disaster mitigation~\cite{leonov2009wearable,boyer1992properties,lv2022smart}. 
Despite these possibilities, the practical deployment of TE devices has faced challenges, primarily due to their typically modest energy conversion efficiency. 
The efficiency of TE materials is assessed through their dimensionless Figure of merit ZT =$\sigma S^{2}T/ \kappa$, a key parameter influencing their overall efficiency. 
The expression for the dimensionless Figure of merit (ZT) is defined as the product of the Seebeck coefficient (S), electrical conductivity ($\sigma$), and absolute temperature (T), divided by the total thermal conductivity ($\kappa$), which encompasses both electronic and lattice contributions. 
The pursuit of identifying thermoelectric (TE) materials with elevated ZT values has been a central focus of research in this field for several decades~\cite{mahan1996best,heremans2008enhancement}.

In thermoelectric cooling devices designed for commercial applications, Bi2Te3 stands out as a commonly employed thermoelectric (TE) material~\cite{hong2018fundamental}, while PbTe is favored for high-temperature TE applications~\cite{gelbstein2005high}. 
Exploration into materials such as NiTiSn and ZrNiSn is ongoing~\cite{downie2014thermoelectric}, driven by their potential in thermoelectric applications. Their remarkable performance can be attributed to the value of ZT near to  1. The efficacy of these materials is primarily influenced by the increased power factor ($\sigma S^{2}T$) or the reduction in thermal conductivity. 
However, owing to the toxic nature of lead-based PbTe and the high cost associated with Tellurium, the production of these thermoelectric (TE) materials is constrained.

 In recent years, the spotlight in thermoelectrics has shifted towards Ag-based chalcopyrites, drawing considerable attention. This increased focus is credited to their exceptional adaptability in crystal structures and electronic band arrangements, as noted in prior research~\cite{cao2020origin,parker2012thermoelectric,rani2024first,jana2025nonempirical}.   The ternary and quaternary Ag-based chalcopyrites such as AgXY$_{2}$ (X=Ga, In and Y=S, Se, Te) and Ag$_{2}$PQR$_{4}$ (P=Mg, Mn, Fe, Zn, Cd, Hg; Q= Si, Ge, Sn; R= S, Se, Te) 
crystallizing in diamond like structure,
 have become a focal point due to their versatile composition and functional adjustability. These attributes position them as promising contenders for applications in thermoelectric devices.
 
Recent investigations have revealed that AgGaTe$_2$ and AgInTe$_2$ boast thermal conductivities of 1.94 and 2.05 W m$^{-1}$ K$^{-1}$, respectively, at standard room temperature, as documented by Charoenphakdee et al. \cite{charoenphakdee2009thermal}. 
This underscores their potential as promising contenders in the realm of thermoelectric materials, given their notably low thermal conductivities. 
Sharma et al. \cite{sharma2014ab} conducted an exhaustive analysis of AgGaTe$_{2}$'s thermal attributes, delving into its Debye temperature, entropy, and heat capacity under diverse pressure and temperature conditions. 
Understanding the intricate interplay among thermoelectric transport characteristics, temperature variations, and carrier concentration is paramount. Such understanding forms the bedrock for discerning the optimal doping carrier concentration, thus propelling further experimental exploration and the advancement of thermoelectric material applications. 
AgGaTe$_2$ emerges as a promising candidate for p-type doping in thermoelectric applications, with an optimal carrier concentration ranging from $10^{19}$ to $10^{20}$ cm$^{-3}$ \cite{parker2012thermoelectric,peng2014enhanced}. Predictions using the PBE+U approach \cite{yang2017prediction} indicate ZT values of 1.38 and 0.91 at 800 K, corresponding to carrier concentrations of $2.12 \times 10^{20}$ cm$^{-3}$ and $1.97 \times 10^{20}$ cm$^{-3}$ for p-type AgGaTe$_{2}$ (AGT) and AgInTe$_{2}$ (AIT), respectively.\\
Moving on to Ag$_{2}$ZnSnS$_4$ (AZTS) and Ag$_{2}$ZnSnSe$_4$ (AZTSe), experimental findings from Li et al. \cite{li2013synthesis} and Kaya et al. \cite{kaya2015synthesis} report band gaps of 1.20 eV and 1.40 eV, respectively, showcasing high absorption coefficients ideal for solar cell applications, as highlighted by Chagarov et al. \cite{chagarov2016ag2znsn}. 
Synthesis efforts led to an impressive 10.7$\%$ efficiency for AZTS \cite{xianfeng2013deposition}. Beyond photovoltaics, AZTS has found utility in photocatalysts \cite{li2013synthesis} and photo-electrochemical applications \cite{yeh2014preparation}. 
Comparisons with Cu$_2$ZnSnS$_4$ (CZTS) reveal similar structural and phase stability but distinct conductivity characteristics, as elucidated by Chen et al. \cite{chen2009crystal} and Chagarov et al. \cite{chagarov2016ag2znsn}. Information about the crystal structure of Ag$_{2}$ZnGeS$_4$ (AZGS) is available in reference ~\cite{PARASYUK201026}. Although no experimental studies have been conducted Ag$_{2}$ZnGeSe$_4$ (AZGSe), theoretical investigations by Nainaa et al. \cite{nainaa2019first} employing mbj-GGA exchange-correlation approximation suggest enhanced absorption and carrier concentration compared to CZTS, suggesting promising prospects for improving solar cell performance.

The thermoelectric properties of silver-based quaternary materials remain largely unexplored. The optimization of parameters like carrier concentration, effective mass, and band gap is essential for maximizing the Figure of merit (ZT), as highlighted by Snyder et al. \cite{snyder2008complex}.
 This Paper employs the multiband Boltzmann transport equations (BTEs) as the analytical framework to delve into and predict the thermoelectric (TE) transport characteristics. The essential parameters, such as the band gap and effective mass pertinent to each band, are meticulously derived through first-principles calculations, serving as pivotal inputs for solving the intricate BTEs. Furthermore, the relaxation time approximation (RTA), predicated upon the multiband carrier transport model elucidated in Ref.~\cite{zhou2010semiclassical}, is judiciously incorporated into the analysis. To showcase the efficacy and versatility of our proposed methodology, an exhaustive investigation is conducted on the TE properties of an array of materials, including AGT, AIT, AZTS, AZTSe, AZGS, and AZGSe. Leveraging the insights garnered from these analyses; we meticulously delineate the optimal carrier concentrations conducive to achieving peak Figure of merit (ZT) for the materials as mentioned above. It is worth noting that this methodology can be seamlessly extended to explore the TE characteristics of various other Ag-based materials similarly, thus offering a comprehensive framework for advancing the understanding and design of high-performance thermoelectric materials~\cite{chung1967voigt}.

 \section{Methods and technical details}

\subsection{Method}
\label{method}
All calculations are performed using the density functional method within screened dielectric dependent hybrid (DDH) by solving a generalized Kohn-Sham (gKS) scheme having the following form of the  screened xc potential ~\cite{WeiGiaRigPas2018}, 
%
{\boldmath
\begin{eqnarray}
\bm{V}_{\bm{xc}}^{\bm{DDH}}(\bm{1},\bm{\epsilon}_{\infty}^{-1};\bm{\mu}) &=& 
\left[ \bm{1} - (\bm{1} - \bm{\epsilon}_{\infty}^{-1}) \, \bm{\mathrm{Erf}}(\bm{\mu} \bm{r}) \right] \bm{V}_x^{\bm{Fock}} \nonumber\\
&-& (\bm{1} - \bm{\epsilon}_{\infty}^{-1}) \bm{V}_x^{\mathrm{sl\text{-}sr},\bm{\mu}} 
+ (\bm{1} - \bm{\epsilon}_{\infty}^{-1}) \bm{V}_x^{\mathrm{sl}} 
+ \bm{V}_c^{\mathrm{sl}}~. \nonumber\\
\label{hy-eq6}
\end{eqnarray}
}
 The key feature of the DDH method is that it takes similarities of the static version of the $GW$, named COH-SEX, where Coulomb hole (COH) is taken care by the GGA approximates and screened exchange (SEX) by Fock term~\cite{CuiWangZhang2018}. 
For reliable DDH calculations using  Eq.~\ref{hy-eq6}, one has to determine the macroscopic static dielectric constant (optical), $\epsilon_\infty$, and screening parameter, $\mu$. Although the PBE calculated $\epsilon_\infty$ often gives good results, self-consistent updating of $\epsilon_\infty$ using DDH is more reliable, especially when PBE predicts the system to be metallic instead of semiconductors~\cite{ghosh2024accurate}. The parameter $\mu$ can also be determined using several procedures. In particular, we consider the procedure proposed in Ref.~\cite{jana2023simple}, which is named as $\mu_{eff}^{fit}$ and obtained using the compressibility sum rule together with linear response time-dependent DFT (LR-TDDFT)~\cite{jana2023simple}. 
\begin{table}[b]
\centering
\caption{%
High-frequency macroscopic static dielectric constants, ion-clamped static (optical) dielectric constants, and electronic dielectric constants ($\epsilon_\infty$) for the PBE and DDH methods, along with screening parameters ($\mu$ in \AA$^{-1}$).}
\label{parameter1}
\begin{tabular}{lccc}
\toprule
Material & $\epsilon_\infty$ (RPA@PBE) & $\epsilon_\infty$ (RPA@DDH) & $\mu = \mu_{eff}^{fit}$ (\AA$^{-1}$) \\ 
\midrule
AGT    & 15.39 & 7.99  & 1.49  \\
AIT    & 11.80 & 7.69  & 1.33  \\
AZTS   & 17.60 & 5.37  & 1.43  \\
AZTSe  & 11.85 & 6.38  & 1.36  \\
AZGS   &  7.02 & 5.35  & 1.39  \\
AZGSe  &  8.45 & 6.55  & 1.47  \\
\bottomrule
\end{tabular}
\end{table}

\subsection{Technical details of calculation procedures}

To optimize the structure and calculate the electronic properties of Ag-based thermoelectric (TE) materials, we  perform first-principles calculations within the DFT as implemented in
Vienna Ab-initio Simulation Package (VASP) code \cite{kresse1996efficient,kresse1996efficiency}. The Perdew-Burke-Ernzerhof form of 
Generalized Gradient Approximation  (GGA-PBE) is used to approximate the exchange-correlation\cite{perdew1996generalized}. 
Handling the complexities of Ag-d electrons poses a challenge, leading us to adopt the dielectric-dependent hybrid (DDH) approach \cite{jana2023simple}. 
This decision is driven by DDH's notable enhancements in the band gaps of Chalcopyrites \cite{ghosh2024accurate}, rendering it well-suited for addressing the unique characteristics of Ag-d electrons. 
This involves calculating the ion-clamped static (optical) dielectric constant, or electronic dielectric constant ($\epsilon_{\infty}$), and the screening parameter ($\mu$). 

A kinetic energy cutoff of $400$ eV 
 is used to limit the number of plane waves in the basis set. 
A Monkhorst-Pack (MP) $\Gamma$-centered $8\times8\times8$ \textbf{k}-points mesh is employed to sample the Brillouin Zone (BZ). 
 The
electronic energies are allowed to converge with
a tolerance of $10^{-6}$ eV across all DFT methods to ensure self-consistency. 
 The structures are relaxed
until  the Hellmann-Feynman forces on atoms diminish to below 0.01 eV/Å$^{-1}$. 
Throughout these computations, we rely on the VASP-recommended Projector-Augmented Wave (PAW) pseudopotentials \cite{kresse1999ultrasoft,woo1997combined}. 
 It is crucial to incorporate
spin-orbit coupling (SOC) in these calculations, given its notable impact on the effective mass tensor, even in materials with low atomic masses \cite{Silva2001}. Elastic constants ($C_{ij}$) are derived from the strain-stress relationship, employing a 15\% strain in the interval of 5\%. According to the Voigt-Reuss-Hill (VRH) theory for a macroscopic system \cite{den1999relation}, corresponding elastic properties such as bulk modulus $B$ and shear modulus $G$ can be evaluated from these constants. The VRH approach is deemed to be in good agreement with experimental measurements. 

The primary focus of the screened DDH approach is to determine the high-frequency macroscopic static dielectric constants. This involves calculating the ion-clamped static (optical) dielectric constant, or electronic dielectric constant ($\epsilon_{\infty}$), and the screening parameter ($\mu$). The steps for performing DDH self-consistent field (scf) calculations are as follows: (i) calculate $\mu_{eff}^{fit}$ as mentioned in Section~\ref{method} with LDA orbitals, (ii) start with $\epsilon_{\infty} (q \rightarrow 0, \omega \rightarrow 0)$ obtained from the PBE functional (RPA@PBE), and plug it into the DDH expression Eq.~\ref{hy-eq6} along with the previously calculated $\mu_{eff}^{fit}$, and (iii) finally, perform the DDH calculation and update $\epsilon_{\infty} (q \rightarrow 0, \omega \rightarrow 0)$ using the result from RPA@DDH, iterating until self-consistency in $\epsilon_{\infty} (q \rightarrow 0, \omega \rightarrow 0)$ is reached.
We report the values of $\epsilon_{\infty}$ (RPA@PBE), $\epsilon_{\infty}$ (RPA@DDH), and $\mu_{eff}^{fit}$ for our systems in Table \ref{parameter1}.

\section{\label{sec:level2}Result and Discussion}

 \subsection{Details of crystal Structure}
 \begin{figure}
    \centering
    \includegraphics[scale = 0.5]{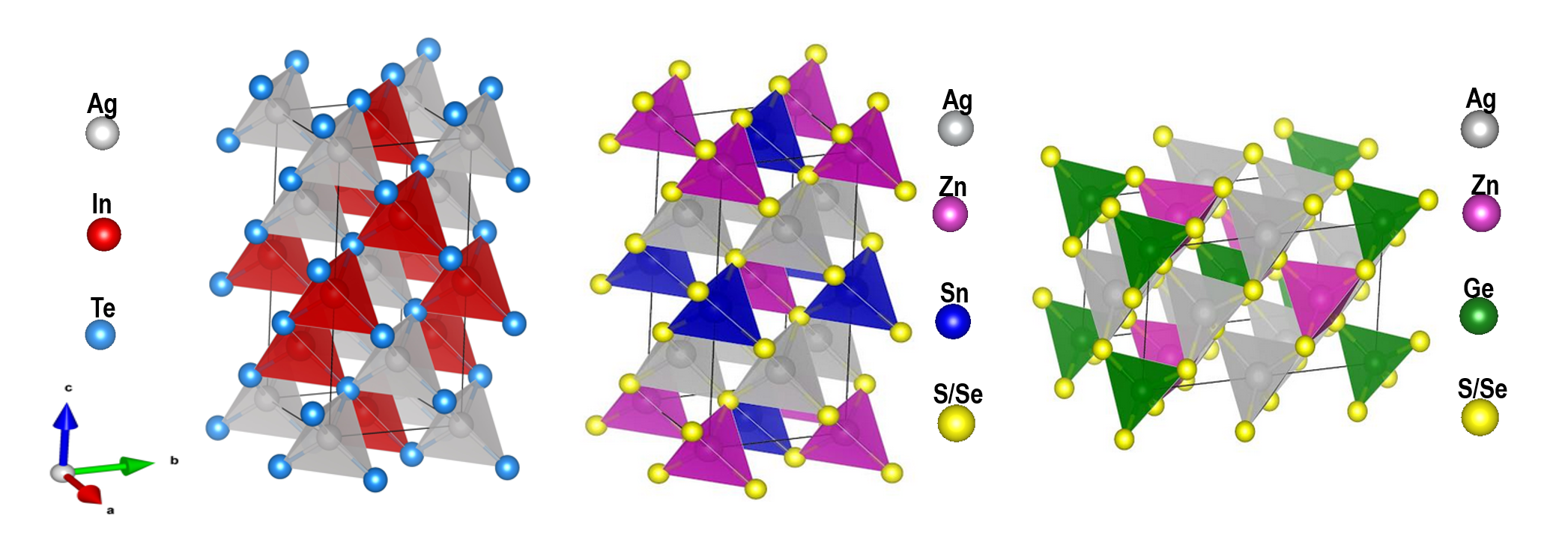}
    \caption{The PBE method was used to optimize the crystal structures of (a) tetragonal AXT (where X = Ga/In) with Te atoms depicted in sky blue, Ga/In atoms in red, and Ag atoms in grey. (b) The crystal structure of AZTQ (where Q=S/Se) shows yellow S/Se atoms, magenta Zn atoms, grey Ag atoms, and blue Sn atoms. (c) In the structure of AZGQ (where Q=S/Se), yellow S/Se atoms are bonded with green Ge atoms, alongside magenta Zn atoms and silver Ag atoms.}
    \label{crys}
\end{figure}
 AXT is experimentally known to crystallize in tetragonal symmetry with space group $I{\bar{4}}2d$~\cite{shaposhnikov2012ab}, while AZSQ and AZGQ (where Q=S/Se) exhibit tetragonal symmetry with  $I{\bar{4}}2m$~\cite{wada2022materials} and $I{\bar{4}}$ space groups. We optimized the lattice constants using PBE before proceeding with further calculations for TE properties. The PBE-optimized lattice constants ($a=b\neq c$) and tetragonal distortion ($\eta$), along with experimental values (when available), are summarized in Table~\ref{lattice}.
 \begin{figure}
    \centering
    \includegraphics[width=\linewidth]{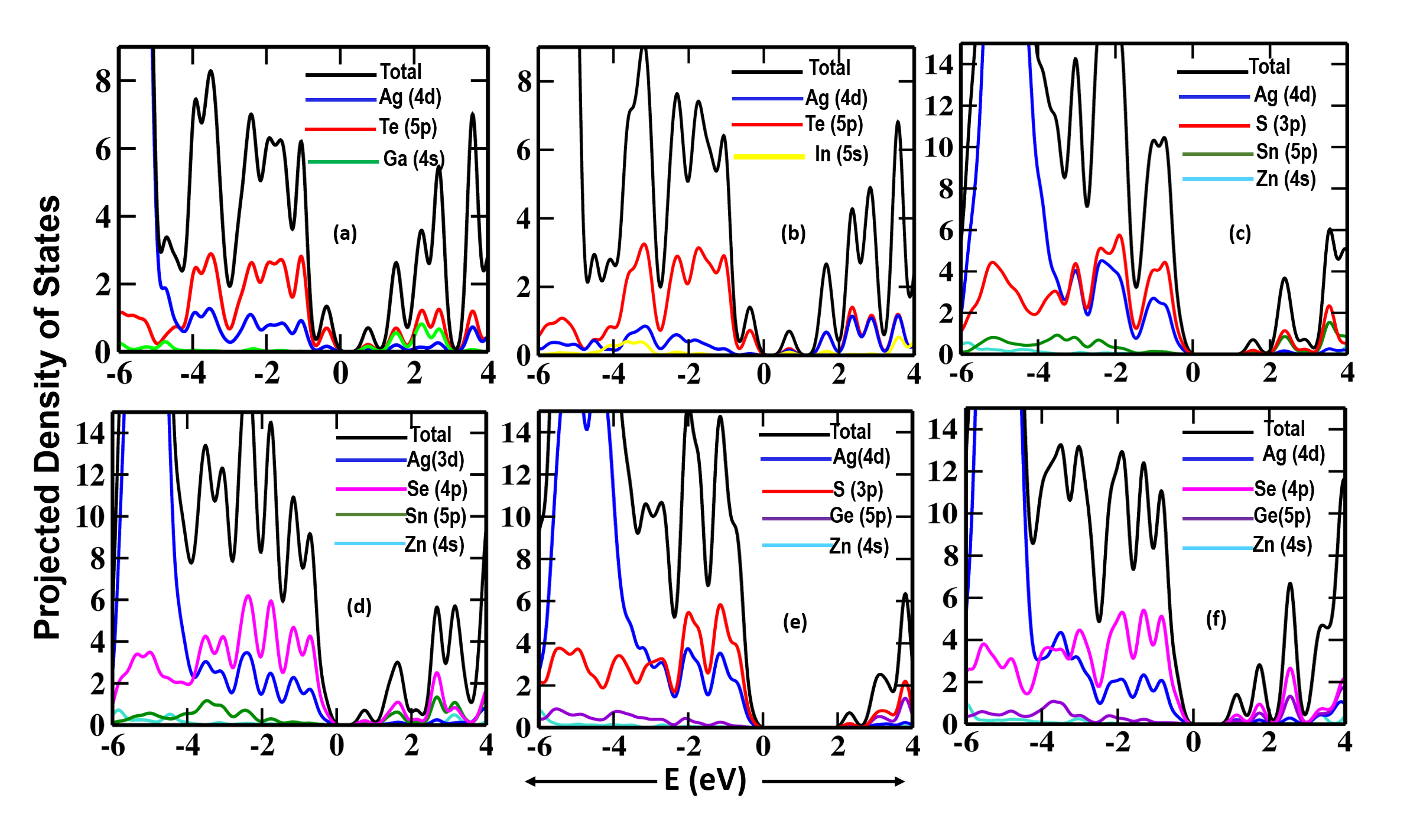}
    \caption{Projected Density of States for (a) AgGaTe$_2$, (b) AgInTe$_2$, (c) Ag$_2$ZnSnS$_4$, (d) Ag$_2$ZnSnSe$_4$, (e) Ag$_2$ZnGeS$_4$, and (f) Ag$_2$ZnSnSe$_4$ from the energy range -6 to 4 eV.}
    \label{dos}
\end{figure}
For PAW pseudopotentials, we selected the valence configurations of the constituent atoms as follows: Ag($4d^{10}5s^{1}$), Zn($3d^{10}4p^{2}$), Sn($4d^{10}5s^{2}5p^{2}$), S($3s^{2}3p^{4}$), Se($4s^{2}4p^{4}$), Ga($3d^{10}4s^{2}4p^{1}$), Ge($3d^{10}4s^{2}4p^{2}$), In($4d^{10}5s^{2}5p^{1}$), and Te($5s^{2}5p^{4}$). The different atomic radii of Ag and the chalcogens (Te, S, and Se) cause local lattice distortions, leading to reduced thermal conductivity. 
These distortions arise from the atomic
size mismatch, which promotes phonon scattering. 
Fig.~\ref{crys} illustrates the crystal structures of AXT (X=Ga/In) and stannite AZPQ (P=Sn/Ge and Q= S/Se), with $a$, $b$, and $c$ representing the lattice constants along the $x$, $y$, and $z$ directions, respectively. 

Fig.~\ref{crys} and Table~\ref{lattice} further illustrate
how the bond formations and distorted environments around Te, S, and Se contribute to this
effect. Fig.~\ref{crys} illustrates the crystal structures of AXT (X=Ga/In) and stannite AZPQ (P=Sn/Ge and Q= S/Se), with $a$, $b$, and $c$ representing the lattice constants along the $x$, $y$, and $z$ directions, respectively. The contribution of individual atomic orbitals to the electronic structure of these materials can be analyzed in detail using Fig.~\ref{dos}, which provides insights into the density of states (DOS) for each atom. This figure illustrates how different orbitals from each atomic species participate in the formation of electronic states across various energy levels, highlighting their role in determining the material's electronic properties.

 \begin{table}[h!]
\centering
\caption{\label{lattice} PBE optimized lattice constants ($a$, $c$), distortion parameter $\eta =c/2a$ used for DDH calculations, and bond lengths of Ag with Te, S, and Se for AXT (X = Ga, In), AZTY, and AZGY with (Y = S, Se) using the DDH and experimental methods (if available). Percent deviations from experimental data (where accessible) are in parentheses.}
\begin{threeparttable}
\begin{tabular}{lcccccccc}
\toprule
 & \multicolumn{2}{c}{$a$ ({\AA})} & \multicolumn{2}{c}{$c$ ({\AA})} & \multicolumn{2}{c}{$\eta$ ({\AA})} & \multicolumn{2}{c}{$d_{Ag-y}$ (y=Te, S, Se)} \\ 
\cmidrule(lr){2-3} \cmidrule(lr){4-5} \cmidrule(lr){6-7} \cmidrule(lr){8-9}
 & PBE & Exp. & PBE & Exp. & PBE & Exp. & PBE & Exp. \\ 
\midrule
\multirow{2}{*}{AGT} 
  & 6.40  & 6.288\tnote{1} & 12.32  & 11.940\tnote{1} & 0.9625 & 0.949\tnote{1} & 2.799  & 2.76\tnote{4} \\ 
  & (1.78) & & (3.1)  &  & (1.4) &  & (1.41) &  \\ 
\midrule
\multirow{2}{*}{AIT} 
  & 6.56  & 6.467\tnote{1} & 13.00  & 12.633\tnote{1} & 0.9908  & 0.977\tnote{1} & 2.811 & 2.78\tnote{4} \\ 
  & (1.43)  &  & (2.9) &  & (1.41)  &  & (1.07) & \\ 
\midrule
\multirow{2}{*}{AZTS} 
  & 5.562 & 5.693\tnote{2} & 12.15  & 11.342\tnote{2} & 1.092  & 0.996\tnote{2} & 2.561  & 2.57\tnote{5} \\ 
  & (2.30)  &  & (7.12)  &  & (9.63)  &  & (-0.3) &  \\ 
\midrule
\multirow{2}{*}{AZTSe} 
  & 5.873 & 5.99\tnote{3}  & 12.606  & 11.47\tnote{3} & 1.07  & 0.960\tnote{3} & 2.656  & 2.65\tnote{6} \\ 
  & (1.95) &  & (9.76) &  & (11.45)  &  & (-0.22) &  \\ 
\midrule
AZGS  
  & 5.76  & --  & 10.633  & --  & 0.24  & --  & 2.578  & --  \\ 
\midrule
AZGSe 
  & 6.03  & --  & 11.25  & --  & 0.24  & --  & 2.678  & --  \\ 
\bottomrule
\end{tabular}
\begin{tablenotes}
    \item[1] Reference~\cite{xiao2011accurate}
    \item[2] Reference~\cite{johan1982pirquitasite}
    \item[3]
    Reference~\cite{wei2017polaronic}
    \item[4] Reference~\cite{cao2020origin}
    \item[5] Reference~\cite{yuan2015engineering}
    \item[6] Reference~\cite{souna2020effect}
\end{tablenotes}
\end{threeparttable}
\end{table}

\subsection{Band Structures}

\begin{figure}[h]
    \centering
    \includegraphics[width=0.5\textwidth]{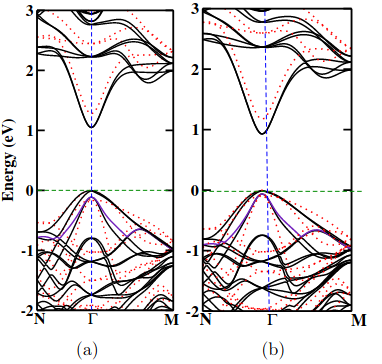}
    \caption{Calculated band structures of (a) AGT and (b) AIT using the dielectric-dependent hybrid (DDH) functional. Bands without spin-orbit coupling (SOC) are shown as dashed red lines, and those with SOC are depicted as solid black lines. The valence band maximum (VBM$_{3,h}$) is highlighted in purple in both cases.}
    \label{bs-1-1}
\end{figure}

We conducted computational investigations to determine the band gaps of AXT (X=Ga, In) and AZPQ (P=Ge/Sn and Q=S/Se) using density functional theory (DFT) with the Dielectric Dependent Hybrid (DDH) functional. Our analysis confirmed direct band gaps at the $\Gamma$ point. We found that the band gaps calculated using the PBE functional significantly underestimated reported experimental values by 80$\%$ to 100$\%$. In contrast, the DDH approach provided band gap values in close agreement with experimental data. 
Additionally, our calculations revealed that the inclusion of spin-orbit coupling (SOC) induced shifts in band positions, consistent with previous studies (Gong et al., 2015).
Table~\ref{tab-all-1} summarizes the band gap values obtained with and without SOC. The importance of band gap in determining thermoelectric properties is underscored by our analysis of the band structures of AgGaTe$_2$ and AgInTe$_2$, as depicted in Fig.~\ref{bs-1-1}  (a) and~\ref{bs-1-1} (b), respectively. We observed significant variations in band gap values when considering SOC, highlighting its influence on electronic properties. Specifically, for AgGaTe$_2$, and AgInTe$_2$ the band gap varied from 1.25 eV to 1.06 eV, and 1.17 eV to 0.94 eV with SOC. The pronounced coupling between Ag-d and Te-p atoms at the valence band maximum (VBM) played a crucial role in determining energy disparities between orbitals.
The substantial spin-orbit coupling (SOC) originating from the heavy Te atoms significantly influences the electronic structure. Due to Te's large atomic number, its p orbitals dominate the valence bands, leading to SOC-induced splitting as shown in figure ~\ref{bs-1-1}. The third-highest valence band shifts lower in energy, while the first two bands converge at the $\Gamma$ point, where SOC effects are weaker. In contrast, the SOC effects from Ag, Ga, and In are smaller due to their lighter atomic masses and less dominant orbital contributions. This behavior is consistent in both AgGaTe$_2$ and AgInTe$_2$, with the primary SOC influence from Te, as shown in Figure 3(a) and 3(b).
Our analysis revealed that the third valence band (VBM$_{3,h}$) significantly influenced thermoelectric properties, as it was situated at -0.105 eV and -0.077 eV for AgGaTe$_2$ and AgInTe$_2$, respectively. A lower band reduces carrier excitation, lowering electrical conductivity, Seebeck coefficient, and electronic thermal conductivity. This generally leads to decreased thermoelectric performance.~\cite{pei2011convergence,zhang2014high}

 Consequently, we find it important to include
all three valence bands in hole transport calculations due to their proximity.

Overall, our findings deepen our understanding of the intricate interplay between band structure, SOC, and thermoelectric behavior, offering valuable insights for materials design aimed at enhancing thermoelectric performance.

\begin{figure}[h]
    \centering
    \includegraphics[width=0.9\textwidth]{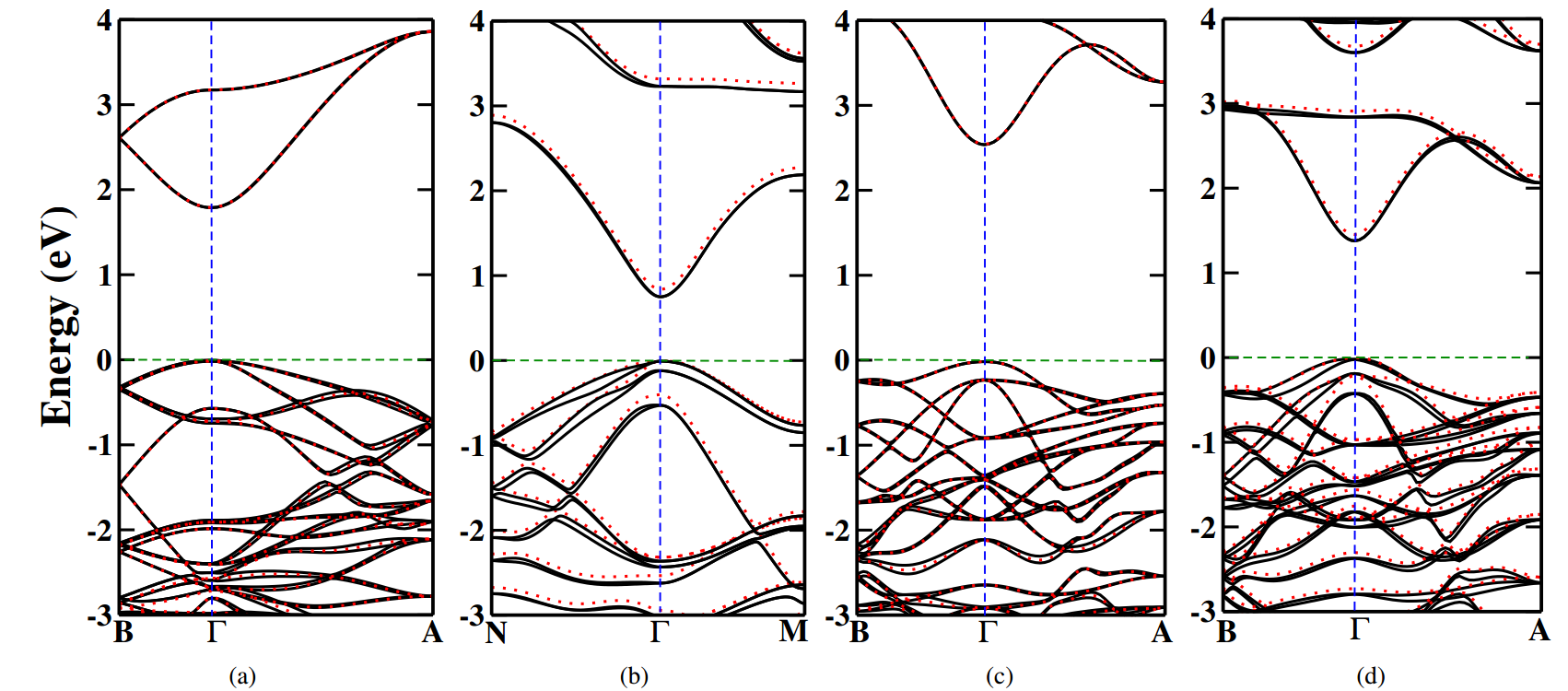}
    \caption{The calculated band structures illustrate the electronic properties of quaternary
chalcopyrite: (a) AZTS, (b) AZTSe, (c) AZGS, and (d) AZGSe. Solid black lines represent
results including Spin-Orbit Coupling (SOC), while red dotted lines depict calculations con-
ducted without SOC. These visualizations offer insights into the intricate energy dispersion
characteristics of these materials, crucial for understanding their potential applications in
transport properties}
    \label{bs-2-1}
\end{figure}

The reported experimental band gaps for AZTS and AZTSe stand at 2.01 eV and 1.4 eV, respectively, as documented in Ref.~\cite{li2013synthesis}. Unfortunately, there are no experimental results available for AZGS and AZGSe. Upon introducing spin-orbit coupling (SOC) in our calculations, noticeable shifts in the band gaps were observed. Specifically, the direct-indirect-direct band gap (DDH) of AZTS, AZTSe, AZGS, and AZGSe transitioned from 1.82 eV, 0.94 eV, 2.56 eV, and 1.45 eV to 1.80 eV, 0.86 eV, 2.55 eV, and 1.40 eV, respectively.

Analysis reveals that sulfide quaternary chalcopyrites exhibited minimal alterations with SOC inclusion compared to selenide Ag-based quaternary chalcopyrites. This can be attributed to the negligible impact of SOC in AZTS and AZGS, primarily due to the lighter mass of sulfur compared to selenium. There is a small SOC splitting in AZTS in valance band compared to AZGS because Sn exhibits stronger SOC than Ge. This trend is visually depicted in Fig.~\ref{bs-2-1}  (a) and~\ref{bs-2-1} (c), showcasing only slight shifts in the top two valence bands of AZTSe and AZGSe, while a distinct shift is evident in the third valence band, as illustrated in Fig.~\ref{bs-2-1} (b) and~\ref{bs-2-1} (d). The corresponding data is accessible in Table~\ref{tab-all-1}. Consequently, these three valence bands play a significant role in hole transport calculations.
\begin{table}[h!]
\centering
\caption{\label{tab-all-1} Calculated band gap with SOC ($E_{g}^{SOC}$), without SOC ($E_{g}^{wSOC}$), and experimental band gap ($E_{g}^{exp}$), along with the valence band maxima (VBM) for the three highest valence bands. Effective masses ($m^{*}$) are given for the lowest conduction band ($e$) and the three highest valence bands ($h_1$, $h_2$, $h_3$), both perpendicular ($\perp$) and parallel ($\parallel$) to the axis in AXT, AZTY, and AZGY using the DDH method. All values are in eV and $m_0$ (free electron mass).}
\begin{threeparttable}
\label{parameter}
\begin{tabular}{lcccccc}
\toprule
& AGT & AIT & AZTS & AZTSe & AZGS & AZGSe \\ 
\midrule
$E_{g}^{SOC}$ (eV)      & 1.06  & 0.94  & 1.80  & 0.86  & 2.55  & 1.40  \\
$E_{g}^{wSOC}$ (eV)     & 1.25  & 1.17  & 1.82  & 0.94  & 2.56  & 1.45  \\
$E_{g}^{exp}$ (eV)      & 1.20\tnote{1}  & 1.04\tnote{2}  & 2.01\tnote{3}  & 1.40\tnote{4}  & --   & --    \\
VBM$_{1,h}$ (eV)        & 0.00  & 0.00  & 0.00  & 0.00  & 0.00  & 0.00  \\
VBM$_{2,h}$ (eV)        & -0.002& -0.017& -0.012& -0.0078& -0.006& -0.0051 \\
VBM$_{3,h}$ (eV)        & -0.105& -0.077& -0.101& -0.018 & -0.166& -0.172 \\
$m^{*}_{1,e\perp}$      & 0.091 & 0.083 & 0.191 & 0.116 & 0.210 & 0.117 \\
$m^{*}_{1,e\parallel}$  & 0.077 & 0.073 & 0.186 & 0.081 & 0.177 & 0.107 \\
$m^{*}_{1,h\perp}$      & 0.256 & 0.445 & 0.679 & 0.625 & 0.765 & 0.525 \\
$m^{*}_{1,h\parallel}$  & 0.085 & 0.083 & 0.770 & 0.172 & 0.215 & 0.123 \\
$m^{*}_{2,h\perp}$      & 0.466 & 0.387 & 0.661 & 0.625 & 0.781 & 0.448 \\
$m^{*}_{2,h\parallel}$  & 0.086 & 0.084 & 0.620 & 0.155 & 0.821 & 0.124 \\
$m^{*}_{3,h\perp}$      & 0.138 & 0.105 & 0.644 & 0.355 & 0.681 & 0.208 \\
$m^{*}_{3,h\parallel}$  & 0.389 & 0.311 & 0.611 & 0.179 & 0.530 & 0.563 \\
\bottomrule
\end{tabular}
\begin{tablenotes}
    \item[1] Reference~\cite{arai2010optical}
    \item[2] Reference~\cite{xiao2011accurate}
    \item[3] Reference~\cite{li2013synthesis}
    \item[4] Reference~\cite{kaya2015synthesis}
   \end{tablenotes}
   \end{threeparttable}
\end{table}

\subsection{Effective mass}

The effective mass \( m^* \) of charge carriers in materials, such as electrons or holes in semiconductors, plays a pivotal role in determining their transport properties. Essentially, it represents how a carrier behaves under the influence of external forces, akin to its inertia. A lighter effective mass implies greater carrier mobility, facilitating faster movement in response to electric fields and leading to higher electrical conductivity. Conversely, heavier effective masses hinder carrier mobility, resulting in lower conductivity~\cite{pei2012low}. The effective mass tensor for electrons and holes is determined using the equation:
\[
m^* = \frac{\hbar^2}{\frac{\partial^2 E(\mathbf{k})}{\partial k^2}} \, ,
\]
where \( E(\mathbf{k}) \) represents the energy dispersion relation, depicting the energy of the particle as a function of its wavevector \( \mathbf{k} \). This tensor can be represented in matrix form as:
\[
\mathbf{M} = \begin{pmatrix}
m_{11} & m_{12} & m_{13} \\
m_{12} & m_{22} & m_{23} \\
m_{13} & m_{23} & m_{33}
\end{pmatrix}
\]
In isotropic materials, such as those with a diagonal effective mass tensor, all off-diagonal elements are zero, retaining components \( m_{11} \), \( m_{22} \), and \( m_{33} \) ~\cite{remsing2020effective}. Here, \( m_{11} \) and \( m_{22} \) represent the transverse effective mass along the \( a \) and \( b \) directions, referred to as \( m^{*}_{i,j}\parallel \), while \( m_{33} \) represents the longitudinal effective mass along the \( c \) direction, denoted as \( m^{*}_{i,j}\perp \).

The effective mass of electrons in the conduction band and holes in the three topmost valence bands is computed along the longitudinal and transverse directions for all six materials, as indicated in Table~\ref{tab-all-1}. This computation involves fitting the band structure data (energy versus \( \mathbf{k} \)) near the \( \Gamma \) point for both parallel and perpendicular directions. The electrons and holes are lighter in ternary compounds than quaternary compounds. The heaviest electron has an effective mass of 0.191 \(m_{0}\) in the perpendicular direction for AZTS quaternary chalcopyrite. In contrast, the heaviest hole for ternary compounds has an effective mass of 0.466 \(m_{0}\), which is lighter than the heaviest hole in quaternary chalcopyrite with an effective mass of 0.821 \(m_{0}\). Next, we will examine the effect of these effective masses on the transport properties of these materials.

\subsection{Electron Transport Properties}
\label{tau-section}

To analyze electronic transport properties, we employ the semi-classical Boltzmann transport equation (BTE) within the relaxation time approximation (RTA)~\cite{zhou2010semiclassical}, using the BoltzTraP2 (BTP2) code~\cite{madsen2018boltztrap2}. Our calculations spanned the temperature range from 600 to 800 K, with 100 K increments, and encompassed hole carrier concentrations up to \(10^{23}\) cm\(^{-3}\). Within the BTE framework, we determine the Seebeck coefficient (\(S\)), electrical conductivity (\(\sigma\)), and electronic contribution to thermal conductivity (\(\kappa_e\)) through the Onsager coefficients, as defined below:

\begin{align}
\bm{S} &= \frac{1}{\bm{q}\bm{T}} \cdot \frac{\bm{L}_{x,y}^{1}}{\bm{L}_{x,y}^{0}},
\label{s} \\
\bm{\sigma} &= \bm{L}_{x,y}^{0},
\label{sigma} \\
\bm{\kappa}_e &= \frac{1}{\bm{q}^2 \bm{T}} \left( \bm{L}_{x,y}^{2} - \frac{(\bm{L}_{x,y}^{1})^2}{\bm{L}_{x,y}^{0}} \right),
\label{kappa}
\end{align}
where \(q\) represents the elementary charge, and \(L_{x,y}^{i}\) denotes the kernel of the generalized transport coefficient at temperature T can be expressed as:~\cite{nag2012electron}
\begin{equation}
\bm{L}_{x,y}^{i} = \bm{q}^2 \sum_{x,y} \bm{\tau}_{x,y} \bm{v}^2_{x,y} \left( \bm{\varepsilon}_{x,y} - \bm{\mu} \right)^i 
\left( -\frac{\partial \bm{f}_{x,y}}{\partial \bm{\varepsilon}_{x,y}} \right),
\label{L}
\end{equation}
with \(i = 0, 1, 2\). Here, \(\tau_{x,y}\), \(v_{x,y}\), \(\varepsilon_{x,y}\), and \(f_{x,y}\) respectively represent the electronic relaxation time, group velocity, energy, and Fermi-Dirac distribution function of the \(x\)th band at wave vector \(y\).
\begin{figure}
    \centering
    \includegraphics[width=0.5\linewidth]{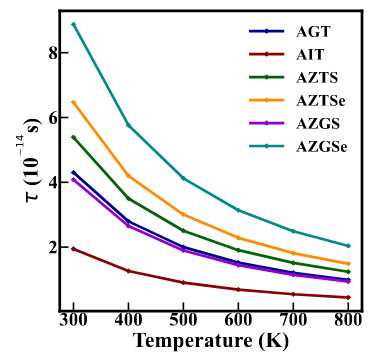}
    \caption{Calculated values of relaxation time  $\tau$ using Deformation Potential theory over the temperature range of 300–800 K.}
    \label{TAU}
\end{figure}
Based on the Eqs.~\ref{sigma},~\ref{kappa} and~\ref{L} calculating the
$\sigma$ and $\kappa_e$ parameters require the relaxation time $\tau$ as an input. This parameter can be calculated using the constant-relaxation-time approximation (CRTA), as outlined in Ref.~\cite{pal2019unraveling, hasan2022first,hao2019design}, the constant $\tau = \tau_0$ will be extracted from the integrals that determine the Onsager coefficients.

In this context, the BTP2 code provides reduced coefficients \(\sigma_0 = \sigma / \tau_0\) and \(\kappa_{e,0} = \kappa_e / \tau_0\), determined by factors such as band structure, temperature, and doping, yet independent of \(\tau_0\). Subsequently, \(ZT\) is derived from these reduced transport coefficients, assuming specific values for the relaxation time \(\tau_0\). The CRTA approach assumes a constant relaxation time \(\tau\) across all energy levels. Consequently, while the power factor relies on \(\tau\), the Figure of merit \(ZT\) remains unaffected by variations in \(\tau\). Recognizing this limitation of CRTA, we opt to employ an alternative method to determine \(\tau\).

To assess the electronic transport characteristics, the deformation potential theory~\cite{bardeen1950deformation,xi2012first} is utilized to analyze the relaxation time. Within the framework of the single parabolic band (SPB) model, the relaxation time for the three-dimensional system is expressed as~\cite{ning2020high,shuai2012theory}:

\begin{equation}
\bm{\tau} =
\frac{\bm{1}}{\bm{3}} \times \bm{2} \sqrt{\bm{2}\bm{\pi}} \cdot 
\frac{\bm{C} \bm{\hbar}^4}{\left( \bm{k}_B \bm{T} \bm{m}_{\text{dos}}^{*} \right)^{\frac{3}{2}}}
\left( \frac{\bm{1}}{\bm{D}} \right)^{2} .
\label{tau}
\end{equation}

Here, \( \hbar \) represents the reduced Planck constant, \( C \) denotes the elastic constant, \( k_{B} \) is the Boltzmann constant, \( m_{\text{dos}^{*}} \) signifies the effective mass of the density of states (as we focus on a p-type system, we consider only the effective mass of holes), and \( D \) refers to the deformation potential energy.\\

\begin{table}[h!]
\centering
\caption{\label{tab-all-4} The calculated single-crystal elastic constants C$_{ij}$ (in GPa), deformation potentials for both electrons D(e) and holes D(h) (in eV), longitudinal (v$_{L}$; in m/s), transverse (v$_{T}$; in m/s), average sound velocity ($\bar{v}$; in m/s), and Debye temperature ($\Theta_{D}$; in K) for materials AXT, AZTY, and AZGY (where X = G, I; Y = S, Se) using the Dielectric Dependent Hybrid approach.}

\begin{tabular}{lcccccc}
\toprule
Property & AGT & AIT & AZTS & AZTSe & AZGS & AZGSe \\
\midrule
C$_{11}$ (GPa)    & 53.90 & 49.55 & 72.11 & 55.99 & 83.27 & 75.97 \\
C$_{12}$ (GPa)    & 30.99 & 31.47 & 58.26 & 43.28 & 51.75 & 48.04 \\
C$_{13}$ (GPa)    & 33.82 & 30.94 & 37.45 & 41.06 & 60.99 & 47.14 \\
C$_{33}$ (GPa)    & 52.72 & 48.24 & 68.01 & 71.26 & 83.22 & 75.97 \\
C$_{44}$ (GPa)    & 41.38 & 35.23 & 25.81 & 41.14 & 51.71 & 47.27 \\
C$_{66}$ (GPa)    & 36.60 & 34.24 & 59.24 & 46.70 & 51.50 & 46.28 \\
D(e) (eV)         & 5.31  & 9.08  & 11.23 & 9.22  & 14.19 & 8.09  \\
D(h) (eV)         & 4.49  & 8.61  & 8.82  & 7.79  & 10.35 & 6.88  \\
v$_{L}$ (m/s)     & 3527.86 & 3373.97 & 3845.77 & 4222.68 & 5217.12 & 4117.47 \\
v$_{T}$ (m/s)     & 2020.93 & 1899.99 & 3895.31 & 2605.64 & 3253.21 & 2248.92 \\
$\bar{v}$ (m/s)   & 2245.14 & 2113.82 & 3878.51 & 2874.30 & 3584.63 & 2508.03 \\
$\Theta_{D}$ (K)  & 211.3   & 192.2   & 403.20  & 284.6   & 380.4   & 253.3   \\
\bottomrule
\end{tabular}
\end{table}
To determine the electronic relaxation time ($\tau$), we computed the single-crystal elastic constants $C_{ij}$ for both ternary and quaternary chalcopyrites. All materials under investigation in this study exhibit a tetragonal structure~\cite{vajeeston2017first}, characterized by six independent elastic constants: $C_{11}$, $C_{12}$, $C_{13}$, $C_{33}$, $C_{44}$, and $C_{66}$, which adhere to mechanical stability criteria given by:

\begin{align}
    C_{11} > C_{12} ,
    \label{eq:1}\\
    2C^{2}_{13} < C_{33}(C_{11} + C_{12})   ,
    \label{eq:2}\\
    C_{44} > 0   ,
    \label{eq:3}\\
    C_{66} > 0    .
    \label{eq:4}
\end{align}

The calculated elastic constants for AXT (where X = Ga, In)and AZPQ (where P=Sn, Ge and Q= S, Se) are presented in Table~\ref{tab-all-4}, satisfying Eqs.~\ref{eq:1}, \ref{eq:2}, \ref{eq:3}, and \ref{eq:4}. The deformation potential ($D$) for the conduction band (electrons) and the valence band (holes) is determined from the strained and unstrained band structures. Subsequently, $\tau$ is evaluated using Eq.~\ref{tau}, and the resulting $\tau$ values are plotted against temperature ranging from $300K$ to $800K$ in Fig.~\ref{TAU}. The relaxation time is expressed in units of $10^{-14}$ s. Notably, The relaxation time decreases with increasing temperature due to enhanced phonon scattering. Among the materials studied, the quaternary chalcopyrite AZGSe exhibits the longest relaxation time, while AIT demonstrates the shortest relaxation time. At 800~K, the relaxation times for the ternary chalcopyrites AGT and AIT are \(0.98 \times 10^{-14}\)~s and \(0.44 \times 10^{-14}\)~s, respectively. In contrast, the quaternary chalcopyrites AZTS, AZTSe, AZGS, and AZGSe exhibit relaxation times of approximately \(1.23 \times 10^{-14}\)~s, \(1.48 \times 10^{-14}\)~s, \(0.95 \times 10^{-14}\)~s, and \(2.03 \times 10^{-14}\)~s, respectively. These values indicate that quaternary chalcopyrites generally have longer relaxation times compared to ternary chalcopyrites. This disparity suggests that quaternary chalcopyrites experience less frequent scattering interactions with lattice defects, phonons, and other charge carriers, resulting in longer relaxation times compared to their ternary counterparts. The deformation potential theory is known to overestimate the relaxation time ($\tau$) due to its simplified treatment of carrier-phonon interactions and neglect of other scattering mechanisms (e.g., polar optical phonons, alloy scattering). In the studied Ag-based materials, these additional scattering mechanisms (particularly in quaternary systems) naturally suppress $\tau$, reducing the degree of overestimation. By drawing comparisons with similar systems, we find that while there may be an overestimation of $\tau$, the relative trends and qualitative conclusions derived from our calculations remain consistent and reliable.

\subsubsection{\textbf{AXT}}
Given the band structure of AXT (where X represents Ga or In) as shown in Fig.~\ref{bs-1-1} (a) and~\ref{bs-1-1} (b), our study facilitates the computation of crucial transport parameters, including the Seebeck coefficient, electrical conductivity, and thermal conductivity, utilizing the Boltzmann Transport Equation (BTE) in conjunction with the Relaxation Time Approximation (RTA). 
Our methodology meticulously incorporates the lowest conduction band and three valence bands proximal to the $\Gamma$ (Gamma) point, while also accounting for the spin degeneracy of each band.

\begin{figure}
    \centering
\includegraphics[scale=0.4]{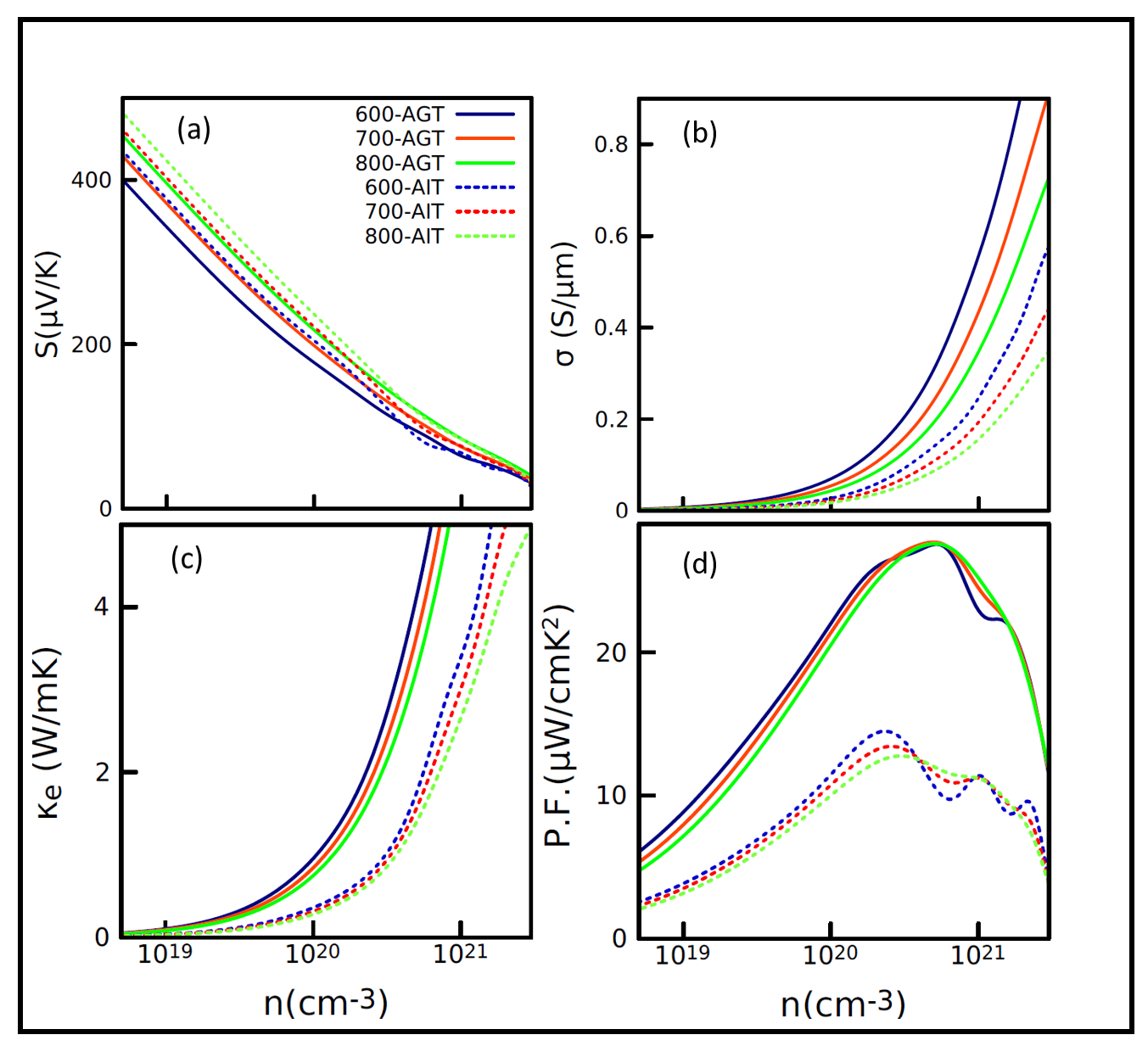}
    \caption{The calculated (a) Seebeck coefficient, (b) electrical conductivity, (c) electron thermal conductivity, and (d) power factor versus carrier concentration are depicted for AGT and AIT at temperatures of 600 K, 700 K, and 800 K.}
    \label{see}
\end{figure}

Equations \eqref{s}, \eqref{sigma}, and \eqref{kappa} illustrate the key parameters. In p-type wide band gap semiconductors, the influence of the bipolar effect is minimal, with hole carriers overwhelmingly dictating transport properties. As temperature rises, relaxation time diminishes due to heightened phonon scattering. Doping emerges as a viable means to enhance electrical conductivity. Our findings align closely with those reported in Yang et al. ~\cite{yang2017prediction}  for AGT and AIT materials. Utilizing the relaxation time ($\tau$) derived from Deformation Potential Theory, as discussed above, we computed the electrical conductivity ($\sigma$) and electrical thermal conductivity ($\kappa_{e}$) in the forms of $\sigma/\tau$ and $\kappa/\tau$, respectively, employing Boltzmann Transport Equation (BTE) theory. At higher temperatures, there's a possibility of minority carrier generation, which can influence the conduction mechanism. This phenomenon, often termed as bipolar thermal conduction, becomes dominant particularly at high temperatures and low carrier concentrations.

Figure~\ref{see} illustrates the dependence of carrier concentration ($n$) on $S$, $\sigma$, S$^2\sigma$, and electronic thermal conductivity ($\kappa_e$). It is evident that $S$ decreases as $n$ increases, whereas $\sigma$ exhibits an opposite trend, ultimately contributing to the enhancement of power factor (S$^2\sigma$). The latter peaks at an optimal carrier concentration before declining. The bipolar effect influences the electronic thermal conductivity ($\kappa_e$) by introducing an additional heat transport contribution from thermally excited minority carriers. Figure ~\ref{see}(c) shows that at low carrier concentrations ($n \sim 10^{19} \, \text{cm}^{-3}$) and high temperatures (800K), the bipolar effect influences $\kappa_e$, resulting in a small but non-zero conductivity due to thermal excitation of minority carriers. As carrier concentration increases above $ 10^{20} \, \text{cm}^{-3}$, the effect diminishes, and $\kappa_e$ rises sharply, dominated by majority carriers. However, the absence of a sign reversal in the Seebeck coefficient ($S$, ~\ref{see}(a)) indicates that the majority carriers remain dominant, limiting the strength of the bipolar effect in the studied regime.  At high temperatures and low carrier concentrations (n), the bipolar effect becomes evident, affecting the electronic thermal conductivity ($\kappa_e$). The peaks in TE coefficients occur at varying carrier concentrations for AGT and AIT conduction, owing to differences in band topology and resulting scattering rates. The power factor (P.F.) for AGT aligns comparably with TE materials from ternary chalcopyrites. The optimal S$^2\sigma$ value of AGT is 27.7 $\mu$W/cmK$^2$ at a hole doping of 5.35 × 10$^{20}$ cm$^{-3}$ at 800 K, as depicted in Fig.~\ref{see}(d), comparable to the P.F. of 25 $\mu$W/cmK$^2$  at an optimal carrier concentration of 8.7 × 10$^{20}$ cm$^{-3}$ for CuGaTe$_2$, belonging to the same category ~\cite{wang2017theoretical}. The significant enhancement in electrical conductivity primarily contributes to these high P.F. values. However, AIT, with an optimal P.F. of 13.53 $\mu$W/cmK$^2$ at a hole concentration of 2.5 × 10$^{20}$ cm$^{-3}$ even at 800 K, falls outside the standard range of TE materials. Nevertheless, its values can potentially be improved through further doping or defect studies.

\begin{figure*}
    \includegraphics[scale=0.42]{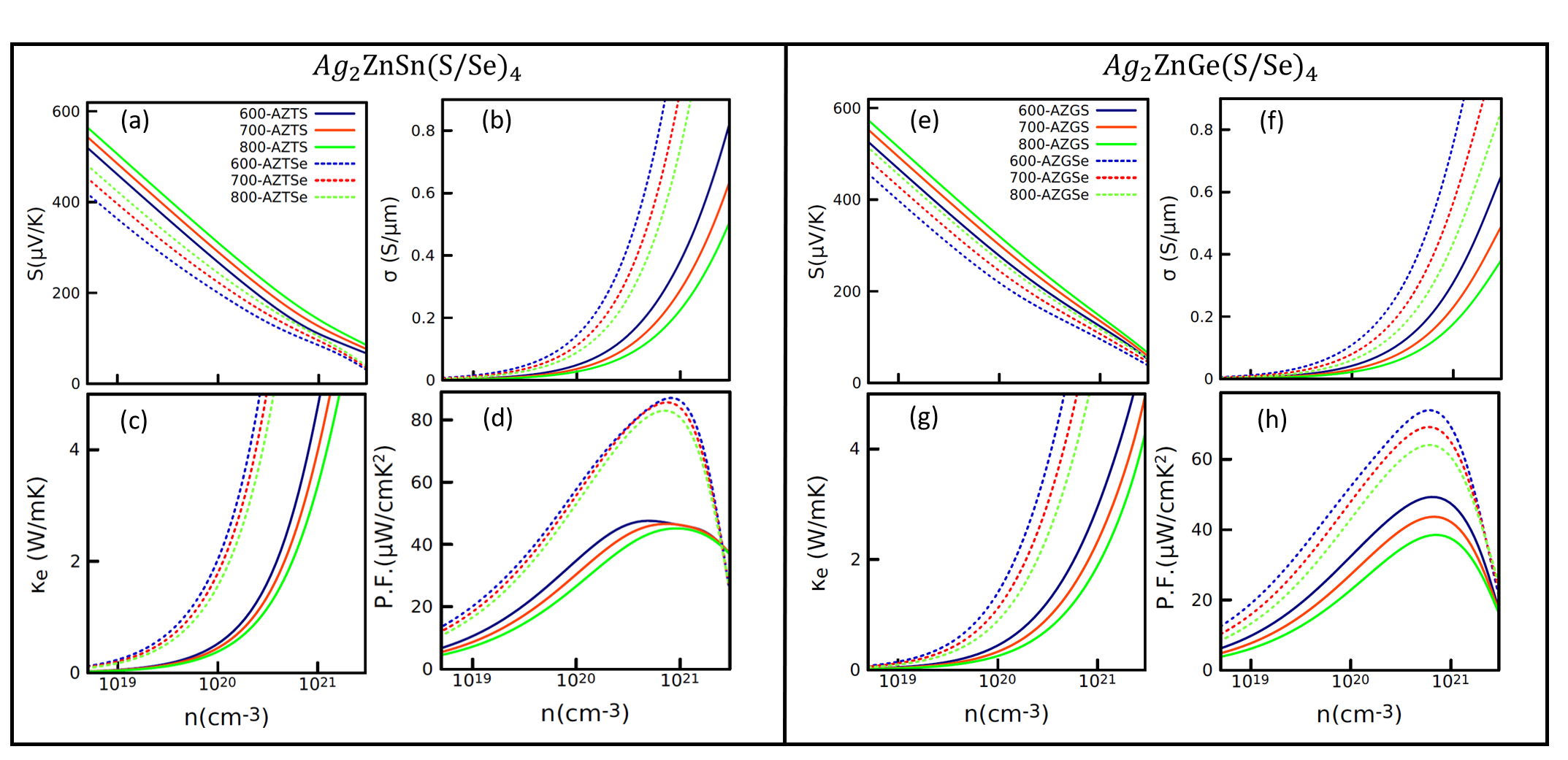}
    \caption{The graph plots the (a)/(e) Seebeck coefficient, (b)/(f) electrical conductivity, (c)/(g) electron thermal conductivity, and (d)/(h) power factor against carrier concentration for Ag$_2$ZnSn(S/Se)$_4$ (on the left) and Ag$_2$ZnGe(S/Se)$_4$ (on the right) across temperatures of 600 K, 700 K, and 800 K.}
    \label{ter-zt}
\end{figure*}

\subsubsection{\textbf{AZPQ}}

Comprehending the transport properties of quaternary chalcopyrites presents a more intricate task compared to their ternary counterparts. This complexity stems from the introduction of an additional element and the resulting intricate interactions within the material ~\cite{li2013synthesis}. Through the analysis of the scattering time ($\tau$), we can extract valuable insights into the mechanisms governing heat and charge transport within these materials ~\cite{hasan2022first}.\\
In AZPQ-type compositions, the incorporation of an extra element relative to ternary chalcopyrites grants finer control over several crucial properties. These include the material's band gap, carrier concentration, and the scattering mechanisms influencing transport. This heightened level of control proves beneficial in applications requiring specific transport characteristics, such as high electrical conductivity for thermoelectric devices or low thermal conductivity for effective heat dissipation.

The Seebeck coefficient ($S$), a crucial parameter in thermoelectric materials, demonstrates a notable enhancement for AZPQ materials as shown in Fig.~\ref{ter-zt}(a) and (e) compared to their AXT counterparts within the same range of hole doping concentration (Fig.~\ref{see}). This observation suggests that AZPQ materials possess a greater ability to convert thermal gradients into electrical voltage due to their superior ability to hold onto charge carriers (higher effective mass or reduced carrier scattering). This is further supported by the significantly higher electrical conductivity ($\sigma$) observed for AZPQ materials compared to AXT counterparts (Fig.~\ref{ter-zt}(b) and (f)). The superior electrical transport properties of AZPQ materials translate to a remarkable improvement in the power factor (PF), a metric that combines $S$ and $\sigma$. As shown in Fig.~\ref{ter-zt}(d), the optimal PF for AZTSe reaches a value of 87.5 $\mu$W/cmK$^2$ at a hole doping concentration of $8.2 \times 10^{20}$ cm$^{-3}$. This value is significantly higher compared to the 47.7 $\mu$W/cmK$^2$ achieved by AZTS material at a lower doping concentration ($5.4 \times 10^{20}$ cm$^{-3}$) in Fig.~\ref{ter-zt}(d).

On the contrary, AZGSe and AZGS, both Ge-based materials, demonstrate a P.F. (S$^{2}$$\sigma$) of 74.7 $\mu$W/cmK$^2$ and 49.4 $\mu$W/cmK$^2$, respectively, at a hole doping level of 6.2 $\times$ 10$^{20}$ cm$^{-3}$ in Fig.~\ref{ter-zt}(h). Comparing Sn and Ge based materials, it's evident that Se-based compounds exhibit higher P.F. compared to S-based ones. If the total thermal conductivity ($\kappa_{e}$ +$\kappa_{l}$) follows a similar trend, favoring Se-based materials, they could outperform S-based materials in terms of TE performance. This is because a higher $\kappa$ facilitates heat flow through the material, reducing the temperature gradient and consequently the voltage generated by the Seebeck effect. In essence, while Se-based materials might excel in voltage generation from a thermal gradient, their potential suffers if their overall thermal conductivity is too high. Optimizing a TE material requires careful consideration of these competing factors to achieve a  maximized ZT and efficient thermoelectric conversion.

\begin{figure}
    \centering
    \includegraphics[width=0.5\linewidth]{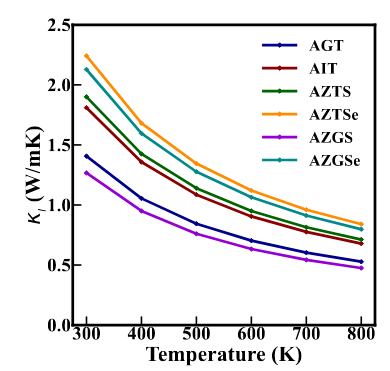}
    \caption{The computed  lattice thermal conductivity obtained through Eq.~\ref{kl}, covering the temperature range from 300K to 800K. }
    \label{tk}
\end{figure}


\subsection{Lattice thermal transport} 

Optimization of the thermoelectric Figure of merit (ZT) requires careful consideration of the thermal conductivity ($\kappa$), which encompasses both electronic ($\kappa_e$) and lattice ($\kappa_l$) contributions. While the Boltzmann transport equation (BTE) framework, detailed in equation (\ref{kappa}), can be used to calculate $\kappa_e$, obtaining $\kappa_l$ via the full linearized phonon BTE is computationally expensive due to the complexity of phonon calculations \cite{ziman2001electrons,li2014shengbte}.
To address this challenge, we can leverage a simpler approach known as Slack's equation to estimate $\kappa_l$ \cite{jia2017lattice,morelli2008intrinsically}. This method offers a practical alternative for materials simulations, particularly when dealing with complex phonon interactions.

\begin{equation}
\bm{\kappa}_{\bm{l}} = \bm{A} \frac{ \bm{\delta} \, \overline{\bm{M}} \, \bm{\Theta}^{3} }{\bm{\gamma}^{2} \, \bm{n}^{2/3} \, \bm{T}} ,
\label{kl}
\end{equation}
 where $\delta^{3}$, $\overline{M}$, $\Theta$, $\gamma$, $n$, and $T$ represent the volume per unit atom, the average atomic mass, the acoustic Debye temperature, the acoustic Grüneisen parameter, the number of atoms in the primitive unit cell, and temperature, respectively. Here, $A$ is a constant given by:

{\boldmath
\begin{equation}
\bm{A} = \frac{\bm{2.43} \times \bm{10}^{\bm{-8}}}{\bm{1} - \frac{\bm{0.514}}{\bm{\gamma}} + \frac{\bm{0.228}}{\bm{\gamma}^{\bm{2}}}} .
\end{equation}
}
Eq.~\ref{kl} is extensively utilized for evaluating lattice thermal conductivity~\cite{morelli2008intrinsically,nielsen2013lone,shinde2006high,bruls2005debye,skoug2010structural}. The Debye temperatures and Grüneisen parameters of acoustic branches can be accurately determined using phonon dispersions obtained from lattice dynamic calculations or experimental measurements. For computing $\Theta$ and $\gamma$, we adopt the approach proposed by Xiao et al.~\cite{xiao2016origin}, which utilizes an efficient formula based on elastic properties, Poisson's ratio $\eta_p$, or sound velocity, to estimate the Grüneisen parameter given by~\cite{sanditov2009gruneisen,jia2017lattice}:

\begin{table}[h!]
\centering
\caption{\label{lattice2} The calculated Voigt, Reuss, and Hill moduli: bulk modulus (B), shear modulus (G), Young’s modulus (E), Poisson's ratio ($\eta_p$), and the ratio of bulk to shear modulus (B/G) for AXT (X = Ga, In), AZTY, and AZGY (Y = S, Se). Values for moduli are in GPa, while ratios are dimensionless.}
\resizebox{0.99\textwidth}{!}{ 
\begin{tabular}{c|ccc|ccc|ccc|ccc|ccc}
\toprule
Material & \multicolumn{3}{c|}{Bulk Modulus (B)} & \multicolumn{3}{c|}{Shear Modulus (G)} & \multicolumn{3}{c|}{Young's Modulus (E)} & \multicolumn{3}{c|}{Poisson's Ratio ($\eta_p$)} & \multicolumn{3}{c}{B/G Ratio} \\ 
\cline{2-4} \cline{5-7} \cline{8-10}  \cline{11-13} \cline{14-16}
 & Voigt & Reuss & Hill & Voigt & Reuss & Hill & Voigt & Reuss & Hill & Voigt & Reuss & Hill & Voigt & Reuss & Hill \\ 
\midrule
AGT & 39.76 & 39.75 & 39.76 & 28.00 & 18.38 & 23.19 & 68.03 & 47.78 & 58.25 & 0.21 & 0.30 & 0.26 & 1.42 & 2.16 & 1.71 \\ 
AIT & 37.12 & 37.10 & 37.11 & 24.56 & 16.22 & 20.39 & 60.37 & 42.47 & 51.70 & 0.23 & 0.30 & 0.27 & 1.51 & 2.28 & 1.82 \\ 
AZTS & 55.01 & 51.50 & 52.35 & 23.44 & 19.05 & 20.43 & 56.7 & 49.99 & 52.87 & 0.27 & 0.33 & 0.31 & 2.34 & 2.70 & 2.56 \\ 
AZTSe & 48.23 & 47.74 & 47.98 & 56.97 & 17.26 & 13.11 & 122.63 & 46.20 & 88.52 & 0.08 & 0.33 & 0.19 & 0.85 & 2.76 & 1.29 \\ 
AZGS & 59.41 & 59.31 & 59.36 & 18.10 & 77.76 & 47.93 & 49.29 & 162.35 & 113.30 & 0.36 & 0.04 & 0.18 & 3.28 & 0.76 & 1.23 \\ 
AZGSe & 55.69 & 55.09 & 55.38 & 33.93 & 20.94 & 27.43 & 84.61 & 55.76 & 70.64 & 0.25 & 0.33 & 0.28 & 1.64 & 2.63 & 4.31 \\ 
\bottomrule
\end{tabular}
} 
\end{table}

{\boldmath
\begin{equation}
\bm{\gamma} = \frac{\bm{3}}{\bm{2}} \Big( \frac{\bm{1} + \bm{\nu}}{\bm{2} - \bm{3} \bm{\nu}} \Big) .
\label{gamma}
\end{equation}
}
The method based on sound velocity for determining the Grüneisen parameter is computationally more feasible than quasiharmonic phonon calculations~\cite{xiao2016origin}. The Poisson's ratio $\eta_p$ is expressed as a function of the shear wave (v$_{S}$) and longitudinal wave (v$_{L}$) sound velocities, respectively:
\begin{equation}
    \eta_p = \frac{1-2(\frac{v_S}{v_L})^{2}}{2-2(\frac{v_S}{v_L})^{2}} .
\end{equation}
The sound velocities v$_{L}$ and v$_{S}$, and the corresponding average velocity ($\overline{v}$), are given by~\cite{kinsler2000fundamentals}:
\begin{equation}
    v_L = \sqrt{\frac{B + 4/3G}{\rho}}  , \quad v_S = \sqrt{\frac{G}{\rho}},
    \quad  \overline{v} = \Big[\frac{1}{3}\Big(\frac{1}{v_{L}^3}+\frac{2}{v_{S}^3}\Big)\Big]^{-1/3}  ,
    \label{velocity}
\end{equation}
where $B$, $G$, and $\rho$ represent the bulk modulus, shear modulus, and density of the compound, respectively. The Debye temperature of acoustic phonons $\Theta$ is given by~\cite{anderson1963simplified}:
\begin{equation}
    \Theta =\frac{h}{k_B}\Big[\frac{3m}{4\pi}\Big]^{1/3} \overline{v}  n^{-1/3}   ,
    \label{theta}
\end{equation}
Here, $h$, $k_B$, and $m$ are the Planck constant, Boltzmann constant, and the number of atoms per volume, respectively, and $n$ represents the number of atoms in the primitive cell. The term $n^{-1/3}$ is used to roughly separate the acoustic branches from the total vibration spectrum~\cite{anderson1963simplified}.

To determine the lattice thermal conductivity ($\kappa_{l}$), which predominantly contributes (98$\%$), Eq.~\ref{kl} was utilized after computing essential mechanical properties of the materials. These properties, including Bulk Modulus ($B$), Shear Modulus ($G$), and Elastic Modulus ($E$), were directly derived from the elastic constants tensors~\cite{vajeeston2017first}. Subsequent calculations involved longitudinal ($v_L$) and transverse ($v_T$) velocities, along with the average velocity $\bar{v}$, as depicted in Table~\ref{tab-all-4} to determine the Poisson's ratio. The values of $B$, $G$, $E$, Poisson ratio ($\eta_{p}$), and Bulk-to-Shear Modulus ratio were determined using the Voigt-Reuss-Hill approximations~\cite{reuss1929berechnung,hill1952elastic} as shown in the Table~\ref{lattice2}. The $B/G$ ratio was assessed to discern the material's brittleness or ductility, with values below 2 indicating ductile behavior.
Once the average velocity was established, the Debye Temperature ($\theta_{D}$) was calculated using Eq.~\ref{theta}, listed in Table~\ref{tab-all-4} for all materials. Subsequently, $\kappa_l$ was evaluated according to Eq.~\ref{kl}. The lattice thermal conductivity, measured in units of W/mK, was then plotted against temperature in Fig.~\ref{tk}). Notably, a decrease in lattice thermal conductivity with increasing temperature was observed.  
 To validate this method's accuracy, our results for AGT and AIT were compared with experimental lattice thermal conductivity values as 1.7 W/mK and 0.94 W/mK at 300K  ~\cite{yusufu2011thermoelectric,zhong2019silver}. Additionally, it was observed that among all materials, quaternary chalcopyrite AZTS exhibited the highest while AZTSe displayed the lowest lattice thermal conductivity. This disparity in lattice thermal conductivities between quaternary chalcopyrites AZTSe and AZGS underscores the intricate physics governing thermal transport phenomena in these materials. The optimized values of $\kappa_l$ were determined to be 0.52, 0.67, 0.71, 0.45, 0.46, and 0.54 W/mK, for AGT, AIT, AZTS, AZTSe, AZGS, AZGSe respectively, at 800K.

\subsection{Figure of merit}
An accurate estimation of ZT requires an accurate calculation of the electron relaxation time ($\tau$) and lattice Thermal conductivity ($\kappa_l$). 

\begin{figure*}
    \centering
    \includegraphics[scale=0.52]{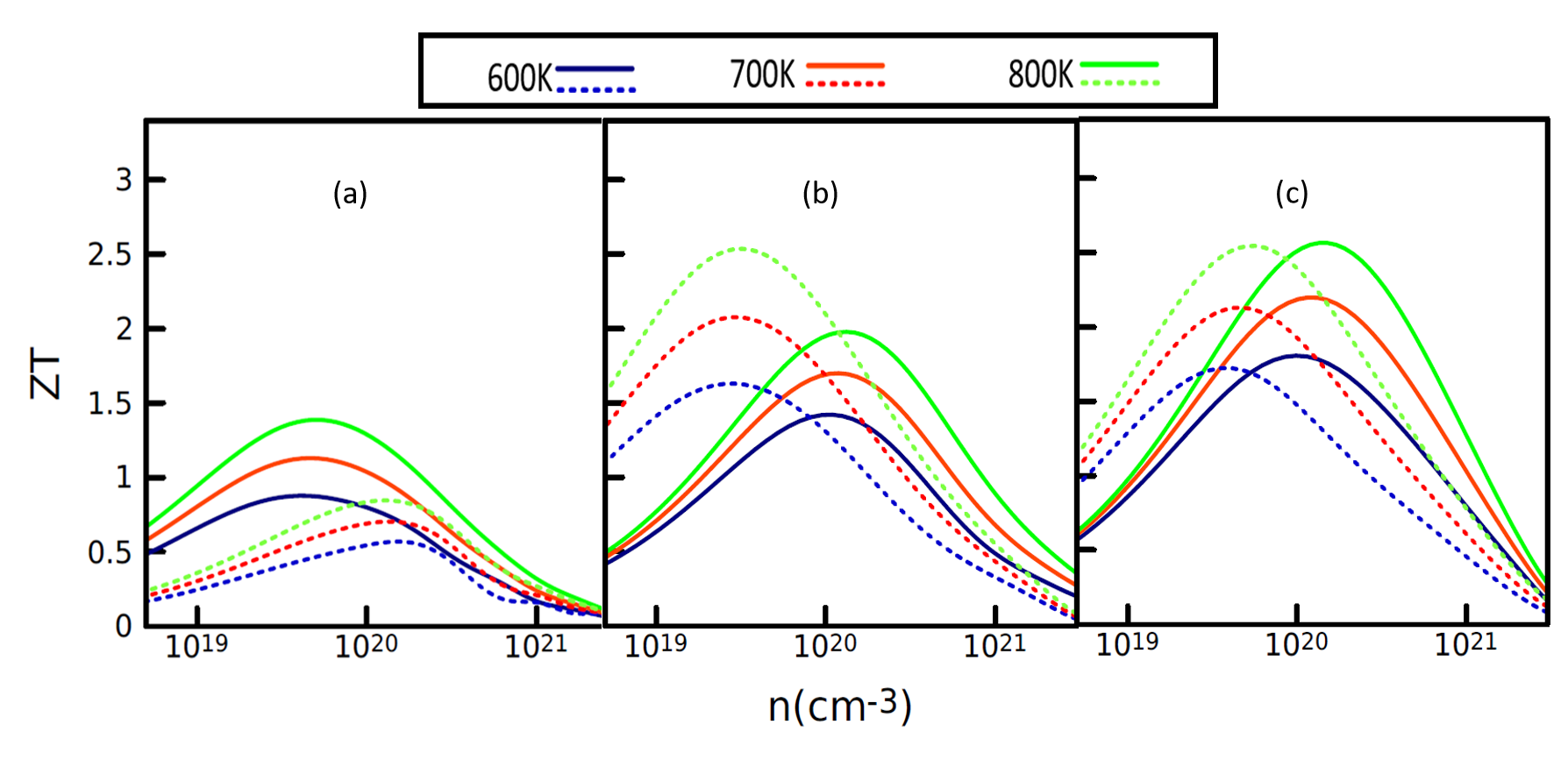}
    \caption{The graphical representation illustrates the computed thermoelectric Figure of merit ZT versus carrier concentration across temperatures of 600 K, 700 K, and 800 K for various compounds:(a) Solid lines depict the ZT values for AGT, contrasting with the representation of ZT for AIT shown by dotted lines. 
(b) Solid lines correspond to AZTS, while dotted lines represent AZTSe. 
(c) The ZT values 
for AZGS and AZGSe are respectively illustrated by solid and dotted lines. }
    \label{zt}
\end{figure*}

The inherent structural complexity of ternary and quaternary chalcopyrites, characterized by low symmetry and large unit cells, poses a formidable challenge for many-body calculations of $\tau$. We employ the deformation potential (DP) method, which accounts for the interactions between charge carriers and acoustic phonons, as initially proposed by Bardeen and Shockley as given in Section~\ref{tau-section}. 
Note that, although the DP approach overlooks Coulombic scattering from polar-optical vibrations, it has been a reasonably good and effective approximation for calculations of parameters of TE materials. 
According to the Fröhlich model ~\cite{huang1988dielectric},  the Coulombic interaction, known as electron-optical-polar scattering, can be effectively screened by using the optical ($\epsilon_{\infty}$) and static ($\epsilon_{s}$) dielectric constants. 
As discussed in the previous work ~\cite{deng2021electronic}, large dielectric constants are prerequisites for efficient TE materials. This requirement  also turns DP into a reasonable approach for calculating $\tau$. By using $\tau$ and $\kappa_l$ we have calculated Figure of merit ZT for all our compounds as shown in Fig.~\ref{zt}. 
Note that the relaxation time decreases with increasing temperature, roughly following a $T^{-3/2}$ relationship as denoted in Eq.~\ref{tau}. Therefore, the variation of ZT with temperature almost resembles a dome-shaped curve. For our system ZT reaches maximum at 800 K. we have plotted the ZT values for our system in the range of 600-800 K. 
For p-type thermoelectric materials, the electronic concentration and electron thermal excitation are obvious at low hole concentration and high temperature. 
We can see in Fig.~\ref{zt}(a-c), at lower hole concentration the difference between ZT values for different temperature is less because of the thermal excitation. 
As we can see, the Figure of merit ZT heavily depends on carrier concentration and temperature. There are few calculations of Figure of merit for AGT and AIT reported in literature. 
~\cite{plata2023harnessing,parker2012thermoelectric,zhong2021computationally} 
Based on our calculations in Fig.~\ref{zt}(a),  AGT exhibits nearly isotropic ZT values of 1.14 and 1.40 at temperatures of 700 K and 800 K, respectively, with a hole carrier concentration of $4.88 \times 10^{19}$ cm$^{-3}$. These ZT values signify the material's ability to efficiently convert heat differentials into electrical energy, making it a promising candidate for high-temperature thermoelectric (TE) applications. 
Conversely, AIT's maximal ZT of 0.86 at a hole doping level of $1.29 \times 10^{20}$ cm$^{-3}$ suggests limitations in its electron transport properties, resulting in less effective thermal energy conversion compared to AGT. 
This failure to achieve a ZT value above 1 underscores AIT's unsuitability for TE devices. Furthermore, at 600 K, AGT's ZT value of 0.90 at a hole doping of $10^{19}$ cm$^{-3}$ indicates suboptimal electron mobility, which diminishes its potential as a TE material in lower-temperature regimes.

In the realm of quaternary chalcopyrites, there have been no experimental validations for our top suggested TE quaternary chalcogenides, particularly those based on Ag. Se-based quaternary chalcopyrites are expected to have a higher Figure of merit compared to S-based ones, according to the power factor (P.F.) comparison between these materials as shown in Fig.~\ref{ter-zt}(d) and (h). For AZTS, the maximum ZT (Fig.~\ref{zt}(b)) of 1.98 at 800 K is achieved with a hole doping of $1.34 \times 10^{20}$ cm$^{-3}$, whereas AZTSe exhibits a hole doping of $3.25 \times 10^{19}$ cm$^{-3}$, resulting in a ZT of 2.53 at 800 K. 
Hole-doped AZGS achieves a nearly isotropic ZT of 2.57 at 800 K. It's worth noting that AZGS, like the majority of other quaternary chalcogenides with CST (cation-substituted tetrahedral) networks, features a quasi-linear conduction band, leading to nearly negligible n-type thermoelectric behavior. AZGSe also falls within the category of p-type thermoelectric materials, achieving ZT values of 1.72 at 600 K, 2.13 at 700 K, and 2.55 at 800 K, with hole concentrations of $4.9 \times 10^{19}$, $5.8 \times 10^{19}$, and $7 \times 10^{19}$ cm$^{-3}$, respectively.  
An intriguing observation is that while the power factor of AZGSe is higher than that of AZGS, as shown in Fig.~\ref{ter-zt}(h), the Figure of merit of AZGS is greater than AZGSe, incorporating the effect of lattice thermal conductivity ($\kappa_l$) as seen in Fig.~\ref{tk}. 
Specifically, AZGS possesses lower lattice thermal conductivity compared to AZGSe. This difference in lattice thermal conductivity contributes to AZGS having a higher Figure of merit than AZGSe, contrary to initial expectations, as shown in Fig.~\ref{zt}(c).
Among the six materials under study, all except AGT and AIT exhibit ZT values greater than 1 at high hole doping concentrations ($10^{19}$–$10^{21}$ cm$^{-3}$) up to 700 K, positioning them as strong candidates for thermoelectric (TE) applications. However, at higher temperatures, such as 800 K, AGT also shows ZT $>$ 1, making it a promising TE material at high temperatures.

 This phenomenon underscores their potential to efficiently convert waste heat into electrical power. However, it's crucial to acknowledge the inherent complexities of real-world materials compared to idealized models based on density functional theory (DFT) calculations. Laboratory-synthesized samples may encounter various scattering mechanisms that can significantly influence their thermoelectric properties, highlighting the need for experimental validation and further investigation.

 Moreover, besides showing
 a high Figure of merit, thermoelectric materials must also exhibit exceptional dopability to tailor their electronic properties effectively. 
 Notably, diamond-like chalcogenides have garnered attention for their outstanding p-type dopability, enabling precise control over band structures and electron behavior. 
 This characteristic holds promise for enhancing thermoelectric efficiencies and advancing sustainable energy conversion technologies, offering a pathway towards practical implementation in diverse applications.

\section{Conclusions}
We have performed first-principles calculations utilizing the Non-empirical Range-separated Dielectric-dependent Hybrid (DDH) approach to study the thermoelectric (TE) properties of ternary and quaternary Ag-based chalcopyrites.  The DDH approach enables us to accurately capture the electronic band structures, crucial for theoretical exploration of thermoelectric (TE) properties in these chalcopyrites. 
Among these materials, AZGS emerges with the highest band gap of 2.52 eV, coupled with substantial spin-orbit coupling effects, primarily attributed to S-atoms. 
These features contribute significantly to favorable electronic transport parameters. 
In our analysis of charge carrier scattering interactions with acoustic phonons, we have incorporated the Deformation Potential Theory to calculate electron relaxation time. 
The resultant values align well with experimental data (wherever available), validating our theoretical framework.  
In this work, lattice thermal conductivity (\( \kappa_L \)) was estimated using Slack’s equation, instead of computationally expensive harmonic phonon calculations. This is a widely accepted approach for semiconductors and insulators where phonon-phonon scattering is the dominant mechanism. While this method assumes phonons primarily behave as quasi-particles and does not fully account for wave-like effects at much higher temperatures, it remains suitable for our study's temperature range. The use of Slack's equation provides a reliable estimate of \( \kappa_L \), and the results align with trends observed in experimental data, validating its application here. Our results suggest that
AZGS can be a very efficient
TE material, with a ZT value of 2.57 at 800 K, surpassing all other studied materials. 
Additionally, AZGSe and AZTSe are found to exhibit promising TE behavior with ZT values of 2.55 and 2.53 at 800 K, respectively. 
AZTS also emerges as a strong TE candidate, with computed ZT of 1.98 at 800 K. 
Overall, the quaternary chalcopyrites exhibit favorable TE performance across the 600-800 K temperature range. Although AGT demonstrates promising TE characteristics at high temperatures, it falls short at lower temperatures.

Our methodology, coupled with the highly accurate DDH exchange-correlation approximation, underscores the suitability of ternary and quaternary Ag-based chalcopyrites
for further applications in thermoelectric devices, offering promising avenues for practical implementation and advancement in sustainable energy conversion technologies.

\section*{Acknowledgments}

The authors gratefully acknowledge the financial support provided by NISER Bhubaneswar. The calculations were performed on the KALINGA and NISER-DFT high-performance computing (HPC) clusters at NISER, Bhubaneswar.

\bibliography{references.bib}
\end{document}